\documentclass[amsfonts,amssymb,amsmath,prl,aps,reprint,notitlepage,twocolumn,superscriptaddress,longbibliography]{revtex4-2}
\usepackage[dvipdfmx]{graphicx}

\usepackage{braket}
\usepackage{here}
\usepackage{comment}
\usepackage{color}
\usepackage{diagbox} 
\usepackage{colortbl}
\usepackage{dcolumn}% Align table columns on decimal point
\usepackage{bbm}
\usepackage{bm}
\usepackage{ulem}%打ち消し線
\usepackage[dvipsnames]{xcolor}

\begin{document}
\preprint{APS/123-QED}
\title{Chirality-induced intrinsic charge rectification in a tellurium-based field-effect transistor}

    \author{Daichi Hirobe}
    \affiliation{Shizuoka University, Suruga, Shizuoka, 422-8529, Japan.}

    \author{Yoji Nabei}
    \affiliation{Institute for Molecular Science, Myodaiji, Okazaki, 444-8585, Japan.}
    \affiliation{The Graduate University for Advanced Studies, Myodaiji, Okazaki, 444-8585, Japan.}
    \author{Hiroshi M. Yamamoto}
    \affiliation{Institute for Molecular Science, Myodaiji, Okazaki, 444-8585, Japan.}
    \affiliation{The Graduate University for Advanced Studies, Myodaiji, Okazaki, 444-8585, Japan.}
\date{\today}

\begin{abstract}
We report gate-induced enhancement of intrinsic charge rectification without p-n junctions in chiral semiconductor Te under magnetic field $B$. 
As gating shifts the chemical potential to the valence band maximum of Te, the charge rectification efficiency is enhanced hundredfold.
By integrating model calculations, we attribute the charge rectification to the $B$-induced asymmetry of the chiral band structure. 
We also show that the carrier density subject to this asymmetry is augmented by a saddle-point structure near the valence band maximum, which further enhances the gate-tunable charge rectification together with its improved switchability and thermal robustness.
\end{abstract}
\maketitle

%\section{Introduction}%%%%%%%%%%%%%%%%%%%%%%%%%%%%%%%%%%%%%%%%%%%%%%%%%%%%%%%%%%%%%%%%%%%%%%%%%%%%%%%
{\it Introduction.}---
Chiral crystal structure is characterized by breaking of mirror and inversion symmetries. This characterization carries over to momentum space and demands that spin degeneracy of the electronic band be lifted by spin-orbit interaction (SOI). As a result, spin-momentum locking emerges and enables one to electrically control electron spins even in nonmagnetic materials. This property in noncentrosymmetric materials has attracted attention in spintronics research~\cite{SpinCurr}. 
Interestingly, the spin $\bm{\sigma}$ can be coupled parallel to the momentum $\bm{k}$ in chiral crystals via SOI~\cite{NakamuraJPSJ2013, OnukiJPSJ2014, YodaSciRep2015, HirayamaPRL2015, ChangNatMater2018}, which has only recently been verified conclusively~\cite{SakanoPRL2020, GattiPRL2020}. 
The resulting spin-momentum locking is hedgehoglike in momentum space and unique to chiral crystals.
The spin-momentum locking is described by chiral SOI, called Weyl-type SOI, in the form of $\sum_i\alpha_ik_i\sigma_i$ with $\alpha_i$ being SOI strengths ($i=x,y,z$).

Chiral materials have been gaining renewed attention as spin polarizers 
since the discovery of chirality-induced spin selectivity (CISS) in chiral molecules~\cite{RayScience1999, NaamanJPCL2012, NaamanNRC2019, EversAdvMat2022}. Recent extension of CISS from molecules to inorganic metals CrNb$_3$S$_6$ and disilicides suggested micrometer-scale spin-propagation lengths by nonlocal transport measurement~\cite{InuiPRL2020, ShiotaPRL2021} and a high spin-polarization rate by SQUID magnetometry~\cite{NabeiAPL2020}.
Additionally, CISS has been integrated into organic-inorganic hybrid materials~\cite{LuSciAdv2019, KimScience2021, BianAdvSci2022, QianNature2022}, in which regularly structured chiral molecules act as multiple spin filters.
These findings could be steps forward to future spintronics, although the mechanism remains to be clarified~\cite{EversAdvMat2022}.
We note that charge transport phenomena related with Weyl-type SOI follow the same selection rule as CISS~\cite{InuiPRL2020, ShiotaPRL2021}. This makes it impossible to distinguish these two effects by symmetry consideration. A criterion for CISS appears to lie in the high efficiency of charge-to-spin conversion beyond conventional spintronic effects. Therefore, it is critical to explore a quantitative approach to transport phenomena caused by spin-momentum locking in chiral crystals.

Elemental Te is the simplest chiral crystal with strong Weyl-type SOI.
Te was recently shown to host a hedgehoglike spin texture near the highly symmetrical points in momentum space by angle-resolved photoemission spectroscopy~\cite{SakanoPRL2020, GattiPRL2020}. 
Additionally, charge rectification caused by spin-momentum locking without p-n junctions~\cite{rikken2001electrical, TokuraNatCom2018} has been reported both for bulk crystals~\cite{RikkenPRB2019} and for nanowires~\cite{CalavalleNatMater2022}.
These studies encouraged us to explore a quantitative approach to charge rectification caused by spin-momentum locking in Te alongside its possible relation with CISS.

In this Letter, we achieve giant charge rectification caused by spin-momentum locking of a p-type thin-film transistor (TFT) of elemental Te.
A high on/off current ratio of $1.7\times10^7$ enables us to reach the valence band maximum (VBM) of Te and to investigate charge transport caused by spin-momentum locking in a special band structure. 
As gating shifts the chemical potential towards the VBM, the charge rectification efficiency is enhanced by 100, which far exceeds a factor of 6 reported for Te nanowires~\cite{CalavalleNatMater2022}.
By integrating TFT characterization, we also reveal that the enhancement starts when the chemical potential approaches a saddle point slightly below the VBM. 
Our observations are rationalized by the semiclassical Boltzmann transport theory which takes into account chiral spin-momentum locking under external magnetic field.
The excellent correspondence indicates that the charge rectification results from the magnetic-field-induced asymmetry of the chiral band structure. The calculation also shows that near-VBM hole carriers contribute dominantly to this effect. 
We suggest that the augmented density of such carriers in the saddle-point band structure improves the on/off ratio of the charge rectification, and makes the rectification robust against thermal fluctuation.

{\it Spin-split energy bands of Te.}---
Elemental Te is a typical chiral semiconductor and the spin-split energy bands have been studied comprehensively~\cite{HirayamaPRL2015, MatibetPSS1969, DoiJPSJ1970}.
The chirality is known to determine the polarity of many properties, including spin-momentum locking~\cite{SakanoPRL2020, GattiPRL2020}, current-induced bulk magnetization~\cite{VorobevPETF1979, ShalyginPSS2012, FurukawaNatCom2017, TsirkinPRB2018, FurukawaPRR2021} possibly protected by the emergent $SU(2)$ symmetry~\cite{RoyArxiv2022}, optical activity~\cite{NomuraPRL1960, FukudaPSS1975, AdesJOSA1975, StolzePSS1977}, second-harmonic generation~\cite{ChengPRB2019}, and diffraction with circularly polarized x rays~\cite{TanakaJPCM2010}.
In elemental Te, the atoms are covalently bound to form spiral chains along the $c$ axis [Fig. 1(a)]. These chains are bound together by van der Waals force in accordance with an enantiomorphic space group, either $P3_121$ or $P3_221$. 
The anisotropic chemical bonding alongside strong SOI yields anisotropic spin textures in momentum space near the {\it H} and {\it H'} points in the Brillouin zone [Fig. 1(b)]. 
As a result, the two upper branches of the valence band near those points consist mainly of $\ket{j_z=+3/2}$ and $\ket{j_z=-3/2}$ with $j_z$ being the $z$ component of total angular momentum of $j=3/2$. Hybridization between these two states yields an energy gap of $\sim0.1$ eV together with a saddle point near the VBM. The top branch of the valence band can be described by the effective Hamiltonian $H_\text{eff}$ given by~\cite{CalavalleNatMater2022}
\begin{align}
    H_\text{eff} =& -{\it\Delta} -C_1\delta k_z^2-C_2(\delta k_x^2 + \delta k_y^2) \nonumber \\
    & + \chi\frac{{\it \Delta}}{\sqrt{\delta k_x^2 + \delta k_y^2}}(\delta k_x\sigma_x + \delta k_y\sigma_y) + \chi\lambda \delta k_z\sigma_z \label{eq:Heffeq},%\\ 
    %& - \frac{g\mu_B}{2}B_z\sigma_z. \nonumber
\end{align}
where $\delta k_i$ $(i=x,y,z)$ represent deviations of $k_i$ from the $H$-point. 
$\chi=+(-)1$ corresponds to the enantiomorphic space group $P3_{2(1)}21$ and the last two terms proportional to $\chi$ represent Weyl-type SOIs.
The parameters were set to those values in Ref.~\onlinecite{CalavalleNatMater2022}.
The higher energy eigenstate of $H_\text{eff}$ is the valence energy band, visualized in Figs. 1(c) and 1(d).
%Our experiment is intended to probe charge rectification near the VBM, as detailed below.
We stress that previous experiments were susceptible to natural hole doping by Te vacancies~\cite{CalavalleNatMater2022}
and failed to probe charge rectification in the vicinity of the VBM.

{\it Experimental details.}---
Thin-film crystals of Te were prepared, washed, and laminated on heavily doped Si substrates covered with a SiO$_2$ dielectric layer [Fig. 1(e)], following a procedure in Ref.~\onlinecite{WanNatElectron2018}.
Te crystals are known to form millimeter-scale monodomains with one handedness~\cite{RikkenPRB2019, SakanoPRL2020} and this monodomain formation carries over to thin-film crystals obtained by the same crystal growth method as ours~\cite{CalavalleNatMater2022}. Therefore, our Te-based TFT was also expected to be of one handedness, although we did not identify the space group of the Te thin-film crystal.
Contact patterns were defined by standard photolithography 
and the contacts were made by depositing Au through a magnetron sputtering method. 
For TFT characterization [dc measurement in Fig. 1(f)], 
we took transfer curves by sourcing a drain voltage along the screw axis 
and measuring a drain current along the same axis. 
For charge rectification measurement [lock-in measurement in Fig. 1(f)], 
we employed phase-sensitive detection of longitudinal voltage 
in a physical property measurement system (Quantum Design). 
An alternating current $I_c(t)=\sqrt{2}I_{\mathrm{rms}}\sin(2\pi ft)$ was sourced with the frequency $f$ = 37.14 Hz and the amplitude $I_{\mathrm{rms}}$ $<100$ nA.
Voltages at first and second harmonics of $f$ were detected by lock-in amplifiers. The longitudinal voltages along the screw axis were measured while external magnetic field $\bm{B}$ was applied along the same direction.
For high-accuracy phase-sensitive detection, the lowest system temperature was limited to 20 K, below which a high sample resistance hindered high-accuracy measurement.

\begin{figure}
\begin{center}
\includegraphics[width=75mm]{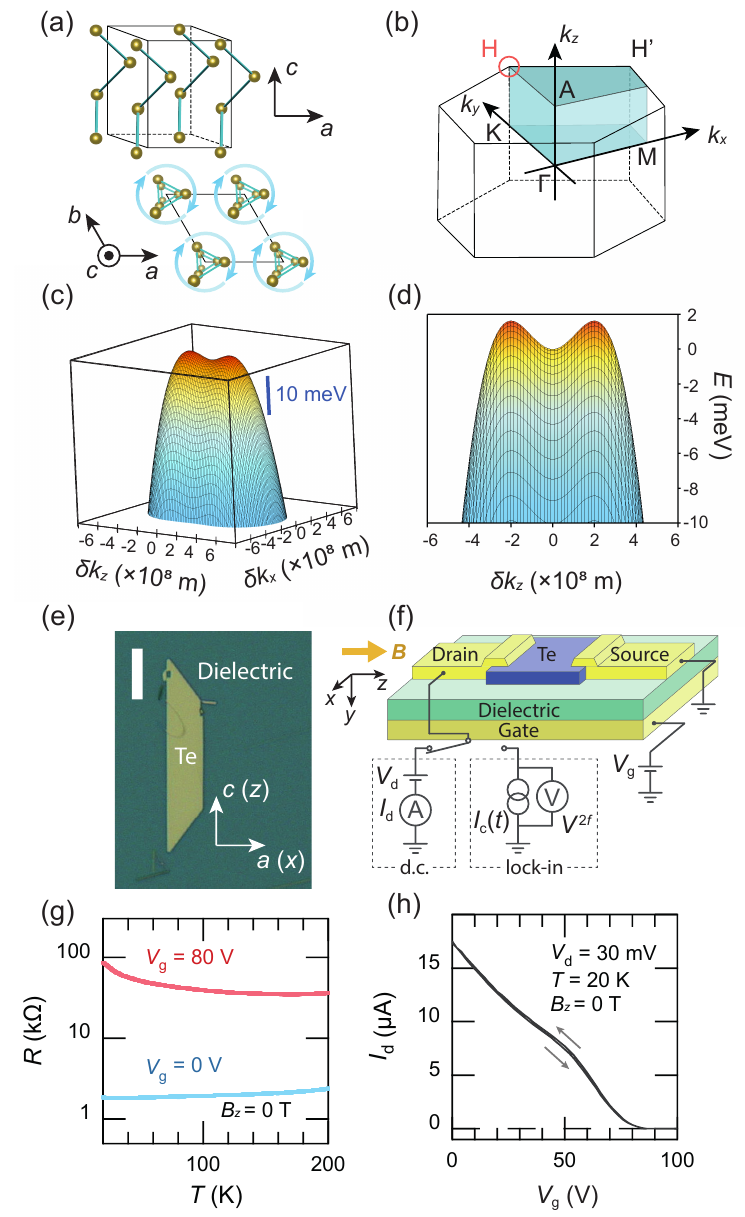}
\end{center}
\caption{
(a) Crystal structure of elemental Te in the space group $P3_221$.
(b) First Brillouin zone of the crystal structure and representative highly symmetrical points. 
$k_{x,y,z}$ represent wave numbers. The valence band maxima are located in the vicinity of the $H$ and $H'$ points.
(c),(d) Valence band in the vicinity of the $H$ point, calculated from the effective Hamiltonian [Eq.~(\ref{eq:Heffeq})]. 
$\delta k_{z,x}$ represent deviations of the corresponding wave numbers from the $H$ point.
(e) Microscopic image of a thin-film crystal of Te 
laminated on the SiO$_2$ dielectric layer. Scale bar 10 $\mu$m.
(f) Schematic of the experimental setup. $V_g$ denotes the gate voltage.
$V_d$ and $I_d$ denote the drain voltage and the drain current in dc measurement 
for TFT characterization. 
$I_c(t)$ and $V^{2f}$ denote the ac excitation and the second-harmonic voltage 
in lock-in measurement for charge rectification. 
$\bm{B} = B_z\bm{e}_z$ denotes the magnetic field applied along the $z$ direction.
(g) Temperature $T$ dependence of the longitudinal resistance $R$ along the $z$ direction, measured with different $V_g$.
(h) $V_g$ dependence of $I_d$, measured at $T=20$ K. $V_d$ was set to 30 mV.
}
\label{fig:tellurium_properties}
\end{figure}

{\it TFT Characterization.}---
In Fig. 1(g), we show the temperature $T$ dependence of the longitudinal resistance $R$ along the $c$ axis, measured at the fundamental frequency. When the gate voltage $V_g$ was set to 0 V, $R$ decreased with decreasing $T$. The metallic conduction indicates that the Te crystal was naturally doped. When $V_g$ was set to +80 V, however, $R$ increased with decreasing $T$ and exhibited a semiconducting property. The result shows that the chemical potential $\mu$ 
was such that valence band states were almost filled up by applying $V_g$,
which changed charge conduction from metallic to semiconducting. 
Because this change took place at positive $V_g$, we confirmed that our Te film was {\it p} doped and that $\mu$ was located in the vicinity of the VBM.
The proximity of $\mu$ to the VBM is also supported by the $V_g$ dependence of the drain current $I_d$. In Fig. 1(h), we show the transfer curve taken at 20 K. 
As $V_g$ is increased from 0 V, $I_d$ decreases and finally becomes indiscernible above $\sim90$ V. The drain-current on/off ratio is $1.7\times10^7$, which is one of the best values reported for p-type Te TFT devices; Te crystals are subject to natural hole doping and do not always exhibit such a high on/off ratio. The excellent TFT property enables us to selectively populate near-VBM hole carriers and to investigate charge transport conveyed by those hole carriers in the presence of spin-momentum locking.

{\it Charge rectification.}---
Having checked near-VBM charge conduction, 
we turn to charge rectification~\cite{rikken2001electrical, TokuraNatCom2018} in the relevant energy range.
Subsequently, we discuss charge rectification in the presence of a uniaxially large Weyl-type SOI, for example,
$\chi\alpha_{z}k_{z}\sigma_{z}$, where $\chi=\pm1$ distinguish left- and right-handed crystals. This SOI represents breaking of inversion and mirror symmetries, namely chirality. As a result,
electrical resistivity has a bilinear correction $\delta\rho(\bm{j},\bm{B})$ with respect to the electric current density $\bm{j}$ and the magnetic field $\bm{B}$. $\delta\rho(\bm{j},\bm{B})$ is given by
\begin{align}
    \delta\rho(\bm{j},\bm{B}) = \rho_0\gamma j_{z}B_{z},
    \label{eq:chiraleq}
\end{align}
where $\rho_0$ is the resistivity at zero magnetic fields. 
This term represents charge rectification and the coefficient tensor $\gamma$ describes the strength as well as the symmetry-adapted selection rule of the charge rectification. 
$\gamma$ due to the anisotropic Weyl-type SOI is finite when $\bm{j}$ and $\bm{B}$ are applied along the $z$ direction. An approximate selection rule was demonstrated for hole transport in Te nanowires~\cite{CalavalleNatMater2022}, probably because of the highly anisotropic Weyl-type SOI in the valence band. 
$\gamma$ is related with chirality via $\gamma \propto \chi$ and the tendency of Te to monodomain formation with one handedness is important to detect $\gamma$.

In the present setup [see also Fig. 1(f)], $\gamma$ appears in an electric field at the second harmonic of $f$, 
$E^{2f} = V^{2f} / l_c$, where $V^{2f}$ is a detected voltage at the second harmonic of $f$, 
and $l_c$ is a distance between the detection electrodes.
$E^{2f}$ is given by~\cite{yokouchi2017electrical}
\begin{align}
    E^{2f} = \gamma \rho^{1f} j_\text{rms}^2 B_z,
    \label{eq:PSDeq}
\end{align}
where $\rho^{1f}$ is the resistivity at the first harmonic of $f$. $E^{2f}$, 
$j_\text{rms}=I_\text{rms}/S_{ab}$ ($S_{ab}$ a cross-sectional area in the $ab$ plane), 
and $B_z$ are all assumed to be along the screw axis ($c$ axis), 
considering the anisotropic Weyl-type SOI.
\begin{figure}
\begin{center}
\includegraphics[width=75mm]{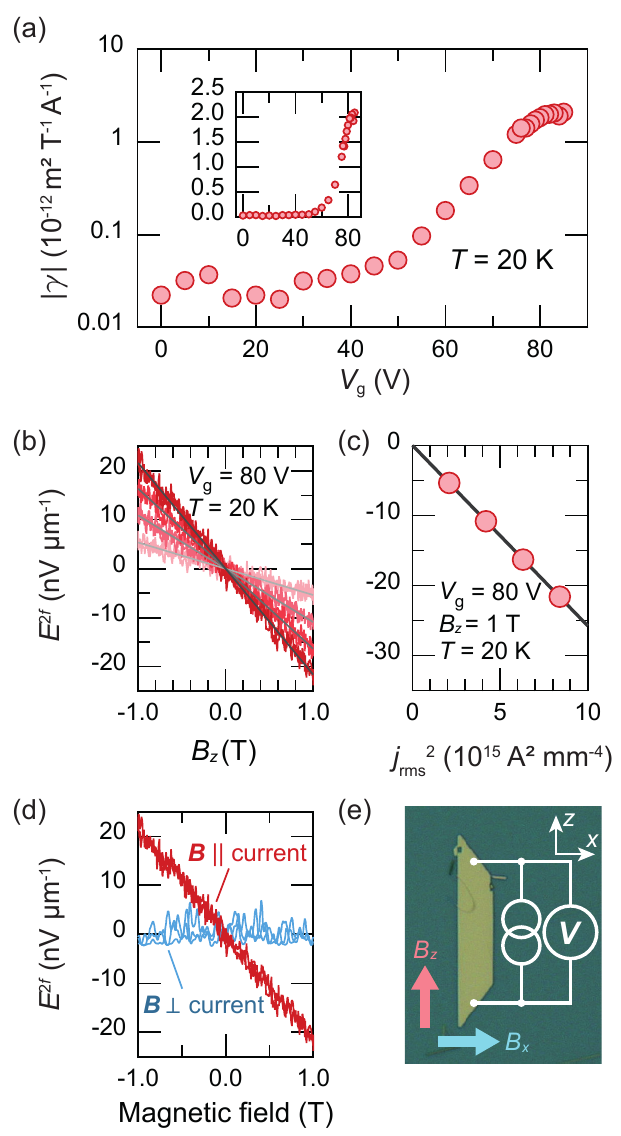}
\end{center}
\caption{
(a) $V_g$ dependence of the charge rectification coefficient $\gamma$ measured at 20 K. The inset shows a linear plot.
(b),(c) $B_z$ (b) and current density $j_\text{rms}$ (c) dependences of the second-harmonic electric field $E^{2f}$. $V_g$ was set to 80 V. $j_\text{rms}$ is increased as the color varies from light to dark in (b) such that $j_\text{rms}^2$ is increased in increments of $2.1\times10^{15} \mathrm{A^2 mm^{-4}}$. 
$E^{2f}$ at $B_z=1$ T is plotted in (c).
(d) Magnetic-field-direction dependence of $E^{2f}$. The direction of $j_\text{rms}$ was fixed along the $c$ axis. 
(e) Schematic of charge rectification measurement overlaid on the microscopic image of the Te thin film. 
$B_z$ and $B_x$ correspond to 
$\bm{B} \parallel \text{current}$ and $\bm{B} \perp \text{current}$ in (d), respectively.
}
\label{fig:gate_dep_gamma}
\end{figure}

In Fig. 2(a), we show the $V_g$ dependence of $\gamma$ at 20 K. 
The corresponding $B_z$ and $j_\text{rms}$ dependences of $E^{2f}$ are shown in Figs. 2(b) and 2(c), 
both of which are consistent with Eq.~(\ref{eq:PSDeq}). 
As $V_g$ is increased from 0 V to 50 V, $\gamma$ increases gradually by a factor of 2. 
However, upon a further increase of $V_g$, $\gamma$ increases exponentially and exhibits a drastic enhancement. 
As a result, $\gamma$ at $V_g = 80$ V is one hundred times larger than at $V_g=0$ V. 
This controllable enhancement is the greatest of three-dimensional normal conductors, 
being second to noncentrosymmetric superconductors~\cite{wakatsuki2017nonreciprocal}.
Additionally, the chiral selection rule of $\gamma$ is supported by 
the magnetic-field-direction dependence of $E^{2f}$ as shown in Fig. 2(d). 
When $\bm{B}$ was applied along the $a$ axis [see also Fig. 2(e)], 
a magnetic-field-linear $E^{2f}$ signal was undetectable. 
If charge rectification resulted from inversion symmetry breaking by $V_g$ along the $b^*$ axis, 
$\gamma$ would be greatest with $\bm{B} \parallel$ $a$ axis and satisfy a selection rule of polar systems~\cite{ideue2017bulk}. Since such a polar selection rule is absent in Fig. 2(d), $V_g$-driven inversion symmetry breaking can be ignored in the present measurement. Instead, our results indicate that the $V_g$-dependent $\gamma$ originates directly from the chiral band structure of Te.

\begin{figure}
\begin{center}
\includegraphics[width=75mm]{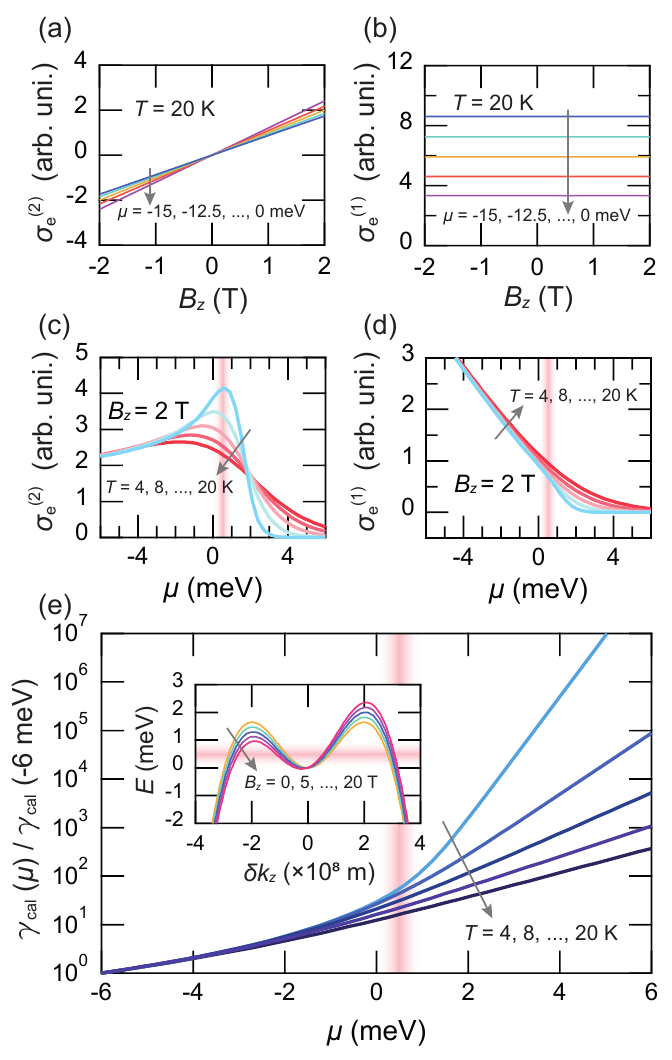}
\end{center}
\caption{
(a),(b) Calculated $B_z$ dependences of the second-order electrical conductivity $\sigma_e^{(2)}$ (a) and the first-order electrical conductivity $\sigma_e^{(1)}$ (b) obtained for each chemical potential $\mu$ in increments of 2.5 meV. $T$ was set to 20 K. 
(c),(d) Calculated $\mu$ dependences of $\sigma_e^{(2)}$ (c) and $\sigma_e^{(1)}$ (d) at several $T$. $B_z$ was set to 2 T.
(e) Calculated $\mu$ dependence of the charge rectification coefficient $\gamma_\text{cal}$, normalized by $\gamma_\text{cal}$ at $\mu=-6$ meV. The inset shows $B_z$-induced asymmetry of the valence band via the Zeeman effect.
}
\label{fig:cal_gamma}
\end{figure}

The enhancement of $\gamma$ near the band edge is reminiscent of that of 
the polar semiconductor BiTeBr~\cite{ideue2017bulk}, 
although the value of $\gamma$ can be one order of magnitude larger in this study. 
The charge rectification in BiTeBr was attributed to $\bm{B}$-induced asymmetry 
of the electronic band via the Zeeman effect alongside its polar atomic arrangement.
Therefore, we calculate $\gamma$, denoted by $\gamma_\text{cal}$ to stress its microscopic nature, 
under the assumption that hole carriers are subject to the Zeeman interaction, 
$H_Z = - \frac{g\mu_B}{2}B_z\sigma_z$, as well as Weyl-type SOI (see also Eq.~\ref{eq:Heffeq}). 
Here $g=\mathcal{O}(1)$ and $\mu_B$ are the effective g-factor and the Bohr magneton, respectively. 
According to Ref. \onlinecite{ideue2017bulk}, $\gamma_\text{cal}$ is given by 
\begin{align}
\gamma_\text{cal} = -\frac{2}{S_{ab}B_z}\cdot\frac{\sigma_e^{(2)}}{\left(\sigma_e^{(1)}\right)^2},
    \label{eq:gamma_cal}
\end{align} 
where $\sigma_{e}^{(1)}$ and $\sigma_{e}^{(2)}$ are first- and second-order electrical conductivities 
with respect to an external electric field, respectively. 
It should be noted that this formalism is independent of a relaxation time $\tau$ 
if $\tau$ is assumed to be constant. 
Therefore, we calculated $\sigma_e^{(1),(2)}$ 
by Boltzmann transport theory with a constant $\tau$, 
using the valence band of the total Hamiltonian $H_\text{eff} + H_Z$. 
$\mu$ and $T$ were taken into account by the Fermi-Dirac distribution function of hole carriers.
In calculating $\sigma_e^{(1),(2)}$, we assumed a two-dimensional limit by setting $\delta k_y$ to 0~\cite{ideue2017bulk}, because $V_g$ was applied along the $b^*$ axis to confine hole carriers in the $ca$ plane~\cite{WanNatElectron2018}.

In Figs. 3(a) and 3(b), we show the calculated $B_z$ dependences of $\sigma_e^{(1),(2)}$ obtained at $T=20$ K. 
$\sigma_e^{(2)}$ is $B$ linear and vanishes at zero fields, consistent with our second-harmonic voltage measurements. 
Interestingly, as $\mu$ is increased towards the VBM at $\mu \sim 1.5$ meV, 
$\sigma_e^{(2)}$ increases and exhibits a peak structure [Fig. 3(c)] 
despite the monotonically decreasing $\sigma_e^{(1)}$ [Fig. 3(d)]. 
When $\mu$ is above the VBM energy, $E_\text{VBM}$, the major $\mu$ dependences of $\sigma_e^{(1),(2)}$ 
stem from the Boltzmann statistics through $e^{-(\mu-E_\text{VBM})/k_{B}T}$ ($k_B$ the Boltzmann constant); 
therefore, $\gamma_\text{cal}$ is enhanced exponentially with the corresponding threshold via 
$\gamma_\text{cal} \propto \frac{\sigma_e^{(2)}}{\left(\sigma_e^{(1)}\right)^2}$, 
as shown in Fig. 3(e). The exponential enhancement with the characteristic threshold reproduces 
the $V_g$-dependent $\gamma$. This demonstrates that near-VBM carriers dominate charge rectification due to the $B_z$-induced band asymmetry [see also the inset to Fig. 3(e)] and that selectively populating those carriers results in efficient charge rectification. 

We recall that the density of states (DOS) near the VBM is augmented by 
the saddle-point structure~\cite{wang2013camel, wang2014camel, peng2014camel} [see also Fig. 1(c)], 
which is absent in noncentrosymmetric semiconductors studied previously. 
Such a DOS contributes to the characteristic peak structures of $\sigma_e^{(2)}$ in Fig. 3(c), 
thereby further enhancing $\gamma$. 
We also suggest that anomalies near the saddle-point structure are more pronounced in experiments than in calculations [compare Figs. 1(h) and 2(a) with Figs. 3(d) and 3(e), respectively]. This indicates that the actual DOS in the saddle-point structure is larger than expected from $H_\text{eff}$ possibly thanks to higher-order corrections to $H_\text{eff}$~\cite{DoiJPSJ1970}. We argue that such a special band structure improves gate switchability of $\gamma$ around $V_g = 50$ V.

In Fig. 4, we compare the $T$ dependence of $\gamma$ with our calculation in which a finite temperature effect was considered via the Fermi-Dirac statistics of hole carriers. We ignored the $T$ dependence of $\mu$ and set $\mu$ to 4 meV in calculation. As $T$ is increased from 20 K to 150 K, the experimental $\gamma$ at $V_g=80$ V ($\mu \sim 4$ meV at 20 K) decreases monotonically. The $T$ dependence is reproduced qualitatively by the calculation, as shown in Fig. 4.
This means that $\gamma$ is due mainly to near-VBM hole carriers and shows that a decreasing ratio of such carriers to the total carrier density is responsible for the decreasing $\gamma$. 

Interestingly, $\gamma$ decreases less rapidly and is more robust against thermal fluctuation than the calculated $\gamma_\text{cal}$. 
Several mechanisms can be responsible for the thermal robustness. 
In one scenario, the result can be attributed to a greater carrier density near the saddle point than expected and likewise the $V_g$-tunable $\gamma$. Such carriers can contribute to $\gamma$  across a wider energy range, thus making $\gamma$ less $T$ dependent. 
Another scenario is a $T$-dependent $\mu$. In $p$-doped semiconductors, $\mu$ shifts upwards with increasing $T$, which helps to selectively populate near-VBM hole carriers.
The last scenario is the relevance of chirality-induced spin selectivity (CISS), which intensifies with increasing $T$~\cite{EversAdvMat2022}. Although CISS-based charge rectification has not been discussed in literature, this problem setting would be worth consideration. 
We note that CISS-induced magnetoresistance (MR) with ferromagnetic electrodes can reach 
tens of percents~\cite{EversAdvMat2022}, which originates from spin-dependent interfacial resistance change. 
This experimental technique was not adopted in the present study; therefore, 
it is difficult to compare the charge rectification efficiency with previous MR values. 
Interfacial MR measurement with ferromagnetic electrodes 
is an important future study to clarify the possible coexistence of CISS in the present system.
Such a MR device may also provide a systematic approach for investigating spin-dependent current--voltage characteristics of CISS without being plagued by the delicate nature of chiral molecular devices.

\begin{figure}
\begin{center}
\includegraphics[width=50mm]{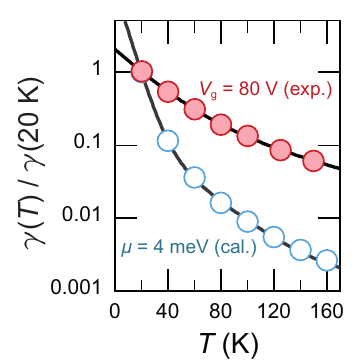}
\end{center}
\caption{
$T$ dependences of the experimental $\gamma$ for $V_g=80$ V and the calculated $\gamma_\text{cal}$ for $\mu=4$ meV, normalized by their respective values at 20 K. Black lines are guides to the eye.
}
\label{fig:temp_dep_gamma}
\end{figure}

{\it Summary.}---
We have demonstrated highly gate-tunable charge rectification caused by spin-momentum locking of a field-effect transistor of elemental Te. 
The charge rectification follows a chiral selection rule and its efficiency is gate variable by two orders of magnitude on demand.
The correspondence with calculations shows that the chiral rectification stems from a concerted effect of Weyl-type spin-orbit interaction and Zeeman interaction with magnetic field.
We expect that Te-based field-effect transistors provide ideal platforms for exploring chirality-induced nonreciprocal phenomena, including CISS.

\begin{acknowledgements}
We thank E. Saitoh for a discussion on the saddle-point structure of elemental Te. 
We also thank T. Sato for experimental assistance.
We are grateful to the Equipment Development Center (the Institute for Molecular Science) for custom-made equipment.
This work was supported by 
Grant-in-Aid for Scientific Research (A) (Grant No. 19H00891) and for Challenging Research (Exploratory) (Grants No. 20K20903 and No. 22K18695) from JSPS KAKENHI, Japan as well as PRESTO “Topological Materials Science for Creation of Innovative Functions" (Grant No. JPMJPR20L9) from JST, Japan. 
A part of this work was performed in Institute for Molecular Science, supported by Advanced Research Infrastructure for Materials and Nanotechnology in Japan (Grant No. JPMXP1222MS0019) of MEXT, Japan.
\end{acknowledgements}

%\bibliographystyle{apsrev4-2}
%\bibliography{ref}

%apsrev4-2.bst 2019-01-14 (MD) hand-edited version of apsrev4-1.bst
%Control: key (0)
%Control: author (72) initials jnrlst
%Control: editor formatted (1) identically to author
%Control: production of article title (-1) disabled
%Control: page (0) single
%Control: year (1) truncated
%Control: production of eprint (0) enabled
%

\end{document}